# Graph-based data-driven discovery of interpretable laws governing corona-induced noise and radio interference for high-voltage transmission lines

Hao Xu[1,2], Yuntian Chen[1,3,*], Chongqing Kang[2,*] and Dongxiao Zhang[1,4,*]

[1] Zhejiang Key Laboratory of Industrial Intelligence and Digital Twin, Eastern Institute of Technology, Ningbo, Zhejiang 315200, China

[2] Department of Electrical Engineering, Tsinghua University, Beijing 100084, P. R. China

[3] Ningbo Institute of Digital Twin, Eastern Institute of Technology, Ningbo, Zhejiang 315200, P. R. China

[4] Institute for Advanced Study, Lingnan University, Tuen Mun, Hong Kong

[*] Corresponding authors.

Email addresses: ychen@eitech.edu.cn (Y. Chen); cqkang@tsinghua.edu.cn (C. Kang); dzhang@eitech.edu.cn (D. Zhang).

**Abstract**

The global shift towards renewable energy necessitates the development of ultrahigh-voltage (UHV) AC transmission to bridge the gap between remote energy sources and urban demand. While UHV grids offer superior capacity and efficiency, their implementation is often hindered by corona-induced audible noise (AN) and radio interference (RI). Since these emissions must meet strict environmental compliance standards, accurate prediction is vital for the large-scale deployment of UHV infrastructure. Existing engineering practices often rely on empirical laws, in which fixed log-linear structures limit accuracy and extrapolation. Herein, we present a monotonicity-constrained graph symbolic discovery framework, Mono-GraphMD, which uncovers compact, interpretable laws for corona-induced AN and RI. The framework provides mechanistic insight into how nonlinear interactions among the surface gradient, bundle number and diameter govern high-field emissions and enables accurate predictions for both corona-cage data and multicountry real UHV lines with up to 16-bundle conductors. Unlike black-box models, the discovered closed-form laws are highly portable and interpretable, allowing for rapid predictions when applied to various scenarios, thereby facilitating the engineering design process.

**Introduction**

High-voltage transmission corridors have become indispensable to the evolving global energy landscape. Large-scale integration of remote renewable resources and the geographical mismatch between generation and demand centres require bulk power transfer over hundreds or even thousands of kilometres (Fig. 1a).[1] Moreover, the environmental footprint of transmission lines is receiving increased attention. A primary concern is the corona effect, which generates audible noise (AN)[2] and radio interference (RI)[3]. These emissions directly impact the local acoustic and electromagnetic environment[4], raising public acceptance concerns and imposing stringent regulatory limits in numerous jurisdictions (Fig. 1a).

The physical mechanisms underlying corona discharge are qualitatively understood and involve the ionization of air molecules and the formation of space charges.[5] However, fast and accurate quantitative prediction of the corona-induced AN and RI remains challenging. The intricate coupling of physical mechanisms has precluded the development of first-principles models that can analytically link these phenomena to basic design parameters such as conductor geometry and surface electric field. Transmission design practices have therefore historically relied on empirically derived formulas (Fig. 1b).[6] Pioneering work by organizations such as the Bonneville Power Administration (BPA)[7] established correlations primarily on the basis of controlled corona cage experiments and line measurements.[8] These methods predominantly employ empirical formulas that express the relationship as a logarithmic function of the surface electric field, conductor diameter, bundle number, and observation distance, with coefficients fitted to data from specific operating conditions.[9,10] While these expressions provide computationally efficient tools, their simplified log-linear forms inherently fail to capture underlying nonlinearities and variable interdependencies[11]. This limitation consequently compromises the predictive accuracy and restricts their generalizability across a wider range of conductor configurations and operating conditions. In recent years, machine learning (ML) has emerged as a powerful alternative for prediction tasks. Techniques such as relevance vector machines[12] and linear mixed-effects models[13] have been employed to refine AN and RI prediction, while principal component analysis combined with neural networks has been used to construct predictive models from relevant features[14]. Although these data-driven models can exhibit superior predictive performance compared with traditional empirical formulas, they operate predominantly as "black boxes." Their internal logic is opaque, offering no intuitive, closed-form mathematical expression that elucidates the underlying relationship between input parameters and corona emissions.[15] The inherent lack of interpretability in black-box models compromises their reliability for critical decision-making and hinders their integration into established design standards (Fig. 1b). Furthermore, these models often operate in isolation from existing physical analysis frameworks, whereas analytical equations can be seamlessly reconciled with theoretical systems to enable broader comparative studies. From a computational perspective, high-dimensional data-driven algorithms entail significant overhead, rendering them impractical for resource-constrained edge computing scenarios. In contrast, explicit mathematical formulas characterized by fewer parameters and minimal computational requirements are highly suitable for low-power, real-time engineering applications. Consequently, there is an urgent need to transform complex, high-dimensional mappings into concise mathematical expressions. Such a formalization bridges the gap between advanced data analytics and practical engineering requirements, facilitating the discovery of symbolic knowledge that is both computationally efficient and physically consistent.

Concurrently, there has been a surge of interest in data-driven equation discovery[16,17], a field aimed at distilling explicit, human-interpretable mathematical expressions directly from observational data. As a representative technique, symbolic regression[18] searches the space of mathematical forms to identify equations that best fit the data without prespecifying the model's structure. By combining neural network-based search with physics-inspired heuristics, advanced algorithms such as AI Feynman[19] and uDSR[20] have been proven to be capable of rediscovering known physical laws and uncovering novel ones directly from observational data. Parallel efforts in differential equation discovery have further shown promise in identifying dynamical system models[21,22]. These advances highlight the transformative potential of AI-driven equation discovery to move beyond mere empirical curve fitting and yield genuine scientific insight. Despite its success across scientific domains[23,24], AI-driven equation discovery encounters hurdles when applied to complex engineering problems. These challenges primarily stem from high dimensionality and the relatively weak, indirect constraints from first principles. For instance, Yang and Park[25] employed evolutionary computation to derive new formulas for predicting audible noise from HVAC lines. However, the inherent limitations of their method restricted the complexity of discoverable terms, ultimately leading to a failure to identify more appropriate nonlinear interactions crucial for accurate prediction. Most symbolic regression algorithms explore the solution space by combining operators and potential variables into expression trees.[26] However, as the number of variables and permissible interaction terms increases, the combinatorial explosion of possible expressions leads to a severe curse of dimensionality, compromising optimization efficiency and stability. Without robust structural constraints, the search often converges to oversimplified local minima or to overly complex, uninterpretable forms that overfit the data. This makes balancing parsimony and accuracy particularly challenging in the presence of measurement noise or limited data coverage. Furthermore, standard objective functions focus predominantly on numerical error and neglect to incorporate qualitative domain knowledge, such as expected monotonic trends or bounded responses under extreme conditions. Consequently, the resulting formulas may achieve numerical accuracy but violate fundamental trends, limiting their practical utility in engineering.

To address these limitations, this work introduces Mono-GraphMD, a graph-based framework for model discovery with monotonicity (Fig. 1c). By representing mathematical expressions as computational graphs, the proposed framework affords greater flexibility in imposing structural constraints, thereby enhancing the efficiency and effectiveness of the search in high-dimensional spaces formed by multiple variables. Furthermore, we integrate domain knowledge directly into the optimization objective by incorporating a monotonicity penalty, ensuring that the discovered formulas adhere to known physical trends within the defined domain and improving their generalizability and reliability. Applying this framework to a dataset from an ultrahigh-voltage corona cage testing facility in China, we discovered novel, interpretable formulas for predicting A-weighted sound pressure level of audible noise ($L_{AN}$) and the radio-interference excitation function (RIEF). These new expressions incorporate previously neglected coupling terms between key variables. A comparative analysis demonstrates that while maintaining comparable complexity in terms of the number of terms, our discovered formulas achieve superior predictive accuracy over established empirical models. The practical validity of the discovered formulas is further corroborated through their ability to predict AN and RI levels along several full-scale UHV alternating current (AC) projects in different countries, where the predictions show excellent agreement with actual field measurements. These results not only validate the effectiveness of the discovered corona emission model but also underscore the broader potential of domain-informed AI-driven model discovery to overcome the limitations of manually derived empirical formulas across science and engineering.

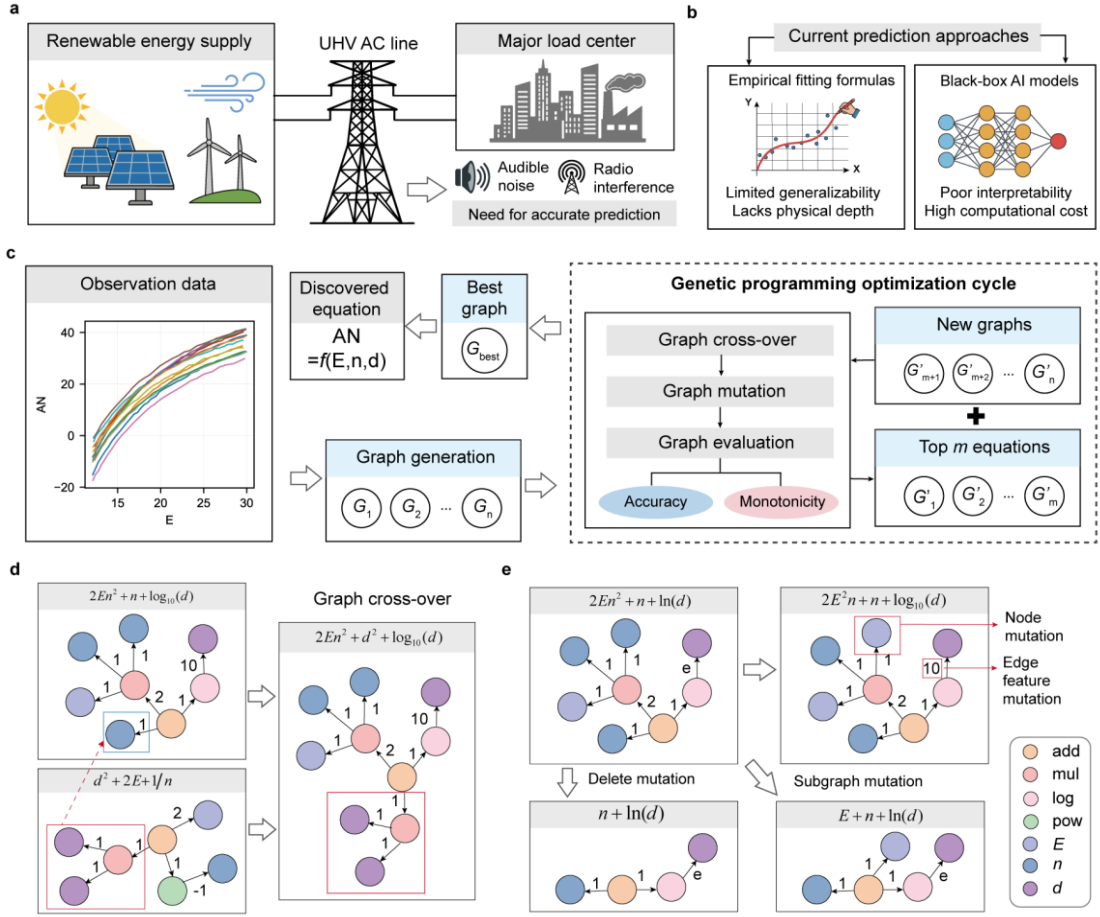

**Fig. 1. Background and overview of the proposed graph-based model discovery framework, Mono-GraphMD. (a)** Schematic illustration of corona-induced audible noise and radio interference in high-voltage transmission lines delivering renewable energy from generation sites to major load centres. **(b)** Summary of key limitations of existing prediction approaches, including empirical models and black-box AI methods. **(c)** Flow chart of the model discovery process, including graph generation, evaluation and optimization. **(d)** An example of graph crossover. **(e)** An example of graph mutation. Here, the equations are expressed as graphs, where nodes represent variables and operators, edges represent computational dependencies, and edge features encode local operations and coefficients.

## Methods

### A monotonicity-constrained framework for discovering physical laws

Mono-GraphMD is a graph-based model discovery framework that incorporates the constraints of monotonicity. Its architecture consists of two tightly coupled components: a graph-based representation of candidate expressions built from a library of graph templates and an evolutionary optimization procedure based on genetic programming that operates directly in the space of expression graphs (Fig. 1c).

**Graph-based representation of equations**

The core idea is to treat each candidate equation as a directed acyclic graph (DAG). A DAG consists of a set of nodes and directed edges. In Mono-GraphMD, nodes represent either operators, such as addition, multiplication, power and logarithm, or variables, while edges encode the computational dependencies between them. Each edge is associated with an edge feature, which is used to parameterize local operations. For example, for a power node (*pow*), the edge feature attached to its incoming edge encodes the exponent; for a logarithm node (*log*), the edge feature specifies the base; for other operators, edge features are used as scalar coefficients that multiply the contribution of the corresponding subgraph. In this way, the structure of an expression is captured by the topology and labels of the graph.

The proposed graph representation deviates from conventional binary expression trees by adopting an *n*-ary connectivity model and an attributed edge framework. While binary trees are constrained by a rigid bifurcation structure that limits operator nodes to a maximum of two children, our architecture allows additive and multiplicative nodes to interface with an arbitrary number of subgraphs. Crucially, this transition enables a significant reduction in topological complexity; many quantities that would otherwise be represented as additional nodes, such as coefficients or operator-specific parameters, are instead encoded as edge attributes attached to the computational relations. This reparameterization reduces the number of explicit nodes and the depth of the topology while separating operand subgraphs from relation-specific attributes. As a result, the children of an operator node are no longer homogeneous, and the operand subgraphs represent the compositional structure, whereas the edge attributes represent operator-coupled parameters that are evaluated in conjunction with the parent operator. This increased flexibility enables compact representations of expressions, which is especially advantageous for problems involving many variables and multiple nonlinear interactions, where tree-based encodings tend to become unnecessarily deep and unbalanced.

Because unconstrained graphs have extremely high structural freedom, a naive exploration of the full graph space would be computationally prohibitive and prone to generating invalid or uninformative expressions. To address this, Mono-GraphMD relies on a graph-template mechanism, which is centred on an additive root node that facilitates the superposition of distinct functional components. This additive structure allows the model to capture global scaling and the cumulative effect of independent features. Within this framework, three primary functional archetypes are stochastically integrated: polynomial subgraphs, rational subgraphs, and logarithmic transformations. The polynomial modules, implemented via a product-of-powers formulation, are designed to explicitly model multivariable interactions, allowing the system to capture nonlinear dependencies between input features. Rational subgraphs introduce reciprocal relationships and normalization effects, while logarithmic templates provide the capacity for scale-invariant transformations and the handling of exponential data distributions. By constraining the search space to these physically meaningful skeletons, the resulting computational graphs maintain a direct correspondence to interpretable mathematical forms. This structured initialization ensures that the discovered equations are not merely numerical fits but are composed of functional building blocks that serve as the basis for future symbolic interpretability and domain-specific analysis.

**Embedding physical monotonicity into equation discovery**

For any fixed graph structure, the numerical coefficients associated with its terms are estimated by regression on the observed data, and the determination coefficient ($R^2$) is calculated to evaluate the numerical accuracy of the expression, which is expressed as $L_{acc}=1-R^2$. Moreover, Mono-GraphMD incorporates monotonicity constraints that encode prior knowledge about the qualitative dependence of the target on each variable. For a given variable $x_j$, a sign $s_j \in [+1,-1]$ is prescribed to indicate whether the function is expected to be globally nondecreasing ($s_j=+1$) or nonincreasing ($s_j=-1$) over a specified domain $D_j$. A finite set of representative values $\{x_j^{(l)}\}_{l=1}^{L} \subset D_j$ is sampled, and all other variables are held fixed at a set of nominal values. The monotonicity loss for variable $j$ is defined as follows:

$$L_{mono,j} = \sum_{l=1}^{L-1} \max\left(0, -s_j[\hat{y}(x_j^{l+1}) - \hat{y}(x_j^l)]^2\right) \tag{1}$$

which penalizes violations of the prescribed monotone trend. The total monotonicity penalty is then $L_{mono} = \sum_j L_{mono,j}$. The overall loss guiding the evolutionary search is given by:

$$L = L_{acc} + \lambda_{mono} L_{mono} \tag{2}$$

where $\lambda_{mono} > 0$ is a weighting parameter. Lower values of $L$ indicate equations that both fit the data well and respect the monotonicity priors over a domain that extends beyond the observed data range. In Mono-GraphMD, sparsity is controlled structurally rather than via explicit regularization, where a maximum number of terms is imposed at the graph level, ensuring that the discovered equations remain compact without the need for additional sparsity penalties.

**Genetic programming optimization in graph space**

Given the graph-based representation, Mono-GraphMD performs structural optimization using a genetic-programming scheme (Fig. 1c). An initial population of equation graphs is generated by sampling from the templates and combining instantiated templates into multiterm expressions. This initial population is then evolved through repeated applications of crossover, mutation and evaluation.

Crossover operates at the level of subgraphs. For two parent graphs, subgraphs attached to the top-level addition node (i.e., individual terms) are randomly selected and swapped (Fig. 1d). In algebraic terms, this corresponds to exchanging one or more terms between two candidate equations. Because each term is itself built from a valid template, the offspring remain within the admissible structural class. Mutation acts in three complementary ways (Fig. 1e): (i) edge feature mutation, in which selected edge features are perturbed; (ii) subgraph mutation, in which an existing subgraph (term) is replaced by a newly sampled template instantiation; and (iii) subgraph addition/removal, in which a subgraph is added to or removed from the set connected to the top-level addition node, subject to a prescribed maximum number of terms. Crucially, all mutations are performed through the same template mechanisms, ensuring that mutated graphs remain valid and do not explode in structural complexity. After crossover and mutation, the population is evaluated and ranked according to a loss function that combines data-fitting accuracy and monotonicity regularization. The top 50% of the graphs are retained as parents for the next generation, while the remainder are replaced by newly generated random graphs, maintaining diversity in the search. This evolutionary loop is iterated until a prescribed number of generations is reached, and the graph with the lowest loss is selected as the final discovered equation.

**Results**

**Discovery of an audible noise predictive formula for alternating current transmission lines**

Under alternating current (AC) excitation, corona discharge initiates on a transmission conductor when the instantaneous surface electric field surpasses the breakdown strength of air. The resulting space charges remain confined to a thin layer near the conductor, as the rapid polarity reversal of the AC field periodically sweeps charges away, preventing the development of extensive space-charge clouds [27]. This confinement implies that the local corona phenomena, including power loss, radio interference, and audible noise, are governed predominantly by local electrical and geometric conditions, with minimal influence from full three-dimensional line configurations.[28] This property enables the corona performance to be characterized by localized testing setups, such as single-phase test lines or corona cages. Consequently, empirical relationships derived from such experiments can be reliably extrapolated to predict the behaviour of transmission lines with arbitrary geometric configurations.

Audible noise data for equation discovery were obtained from corona-cage experiments involving thirteen bundle-conductor configurations.[29] The test matrix covered 8-, 6- and 4- subconductor bundles composed of four subconductor types, namely, LGJ630/45, LGJ500/35, LGJ400/35 and LGJ300/50. For each configuration, the conductor bundle was energized at different voltage levels to realize a wide range of surface electric gradients. A precision noise analyser recorded the emitted sound and converted it into A-weighted sound pressure levels $L_{AN}$ referenced to μW/m, which are defined as follows:

$$L_{AN}^{(\mu W/m)} = 10 \log_{10}\left(\frac{P_{AN}}{1\mu W/m}\right) \tag{3}$$

where $P_{AN}$ is the measured A-weighted acoustic power per unit length. In some literature, $L_{AN}$ referenced to pW/m is used, and the conversion between the two is expressed as follows:

$$L_{AN}^{(\mu W/m)} = L_{AN}^{(pW/m)} - 60 \tag{4}$$

Informed by prior evidence[29] that bundle spacing exerts only a minor influence on audible noise within standard design ranges, this parameter was consequently excluded to prevent unnecessary model complexity. The input variables for equation discovery are thus the surface gradient ($E$, kV/cm), subconductor count ($n$), and subconductor diameter ($d$, cm), with the A-weighted sound pressure level ($L_{AN}$) as the target variable.

The graph-based equation-discovery framework was configured with an operator set ($+, \times, pow, log$), a population size of 500, and 200 optimization iterations. Although this operator set is small, it is functionally expressive, since subtraction can be implemented as addition with a negative multiplicative coefficient, and division is represented as multiplication by an inverse power. A monotonicity penalty with a weight of 0.01 was incorporated into the objective function. For the present application, $L_{AN}$ is expected to increase monotonically with $E$, $n$ and $d$ over the design-relevant range. These expectations were encoded by enforcing global monotonicity with respect to each of the three variables. To examine the trade-off between accuracy and structural simplicity, the maximum number of terms in the candidate formulas was restricted to 3, 4 and 5 (Fig. 2a). The best three-term, four-term and five-term expressions obtained under these settings are written as follows:

$$L_{AN} = 0.0878En + 72.3\log(d) - \frac{648.7}{E\log(E)} \tag{5}$$

$$L_{AN} = 0.093En + 55.02\log(d) - \frac{591}{Ed^2} - \frac{5448}{E^2} \tag{6}$$

$$L_{AN} = 0.0116n^2 d - \frac{102.4n}{E\ln(E)+d^2} + 9.216d + 19.13\ln(n) - \frac{677.3}{E} \tag{7}$$

For comparison, a conventional polynomial-type regression model with a structure similar to that of classical formulas was also fitted to the same data:

$$L_{AN} = 1.022n + 10.4d + 30.839 - \frac{933.633}{E} \tag{8}$$

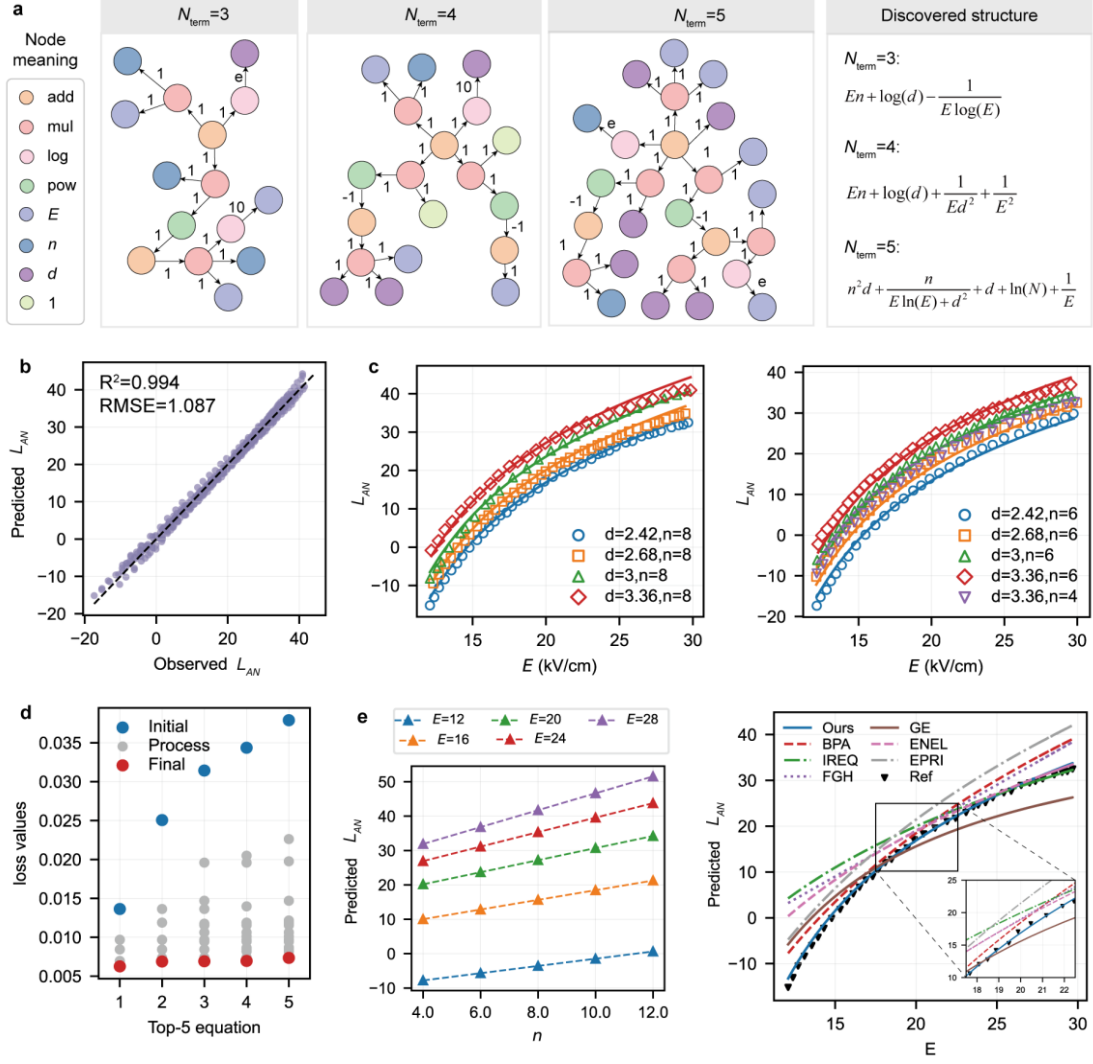

**Fig. 2. Results of data-driven equation discovery for the A-weighted sound pressure levels $L_{AN}$. (a)** The discovered best graphs and formula structures under different numbers of maximum terms. **(b)** Comparison between the predicted and observed $L_{AN}$ across all data. **(c)** Comparison between the predicted and observed $L_{AN}$ for subconductor counts $n$=8, 6, and 4 with different subconductor diameters $d$. Here, the scatters represent the observed data, and the lines represent the predictions from the discovered formula. **(d)** Loss values of the top-5 equations during optimization. **(e)** The predicted $L_{AN}$ under different subconductor counts $n$ and surface gradients $E$. **(f)** Comparison between the discovered formula and classical empirical models for the 8×LGJ300/50 bundle conductor.

Across the whole dataset, the three discovered formulas achieved root-mean-square errors (RMSEs) of 1.087, 1.033 and 0.917, with mean relative errors (MREs) of 0.297, 0.308, and 0.248 for Eqs. (5)–(7), respectively, whereas the polynomial baseline in Eq. (8) yielded an RMSE of 1.714 and an MRE of 0.336. Thus, even the simplest three-term discovered law reduces the RMSE by approximately 40% relative to a carefully tuned polynomial expression. The mathematical structures of Eqs. (5)–(7) provide insight into the underlying dependencies. The presence of linear and log-linear components in the discovered equations mirrors the forms used in traditional empirical laws. In Eq. (5), the nonlinear rational term $\frac{1}{E\log(E)}$ effectively captures the observed saturation of audible noise at high field gradients, which is difficult to identify using conventional equation discovery methods. Here, Eq. (5) is selected for subsequent prediction and analysis. Although Eq. (7) achieves the lowest RMSE, its marginal improvement over the more compact Eq. (5) does not justify the added complexity of its functional form. Therefore, Eq. (5) is favoured because it offers an optimal balance between accuracy and interpretability for practical application. The predictive performance is illustrated in Fig. 2b and c. The predicted curves almost coincide with the experimental markers, capturing the nonlinear increase in $L_{AN}$ with $E$ and the systematic upwards shift of the curves with increasing $d$. This agreement confirms that the discovered formulation correctly captures the dependence of audible noise on the underlying variables. Moreover, the optimization traces showed a steady decrease in the loss of the top-5 equations, indicating that the graph-based search converged reliably under the monotonicity constraints (Fig. 2d). The monotonic behaviour of the discovered law is evident in Fig. 2e, where the predicted curves increase smoothly with the variable $E$ across the entire domain and exhibit no local decreases or unphysical oscillations.

**Comparison between the discovered audible noise predictive formula and existing empirical models**

The performance of the discovered formula is then benchmarked against that of classical empirical models using both corona-cage experimental data and field measurements from UHV AC transmission lines. The predicted audible noise level $L_{AN}$ as a function of surface gradient $E$ for a representative conductor configuration is compared between the discovered formula and widely used empirical formulas, including BPA, IREQ, FGH, GE, ENEL and EPRI (Fig. 2f). The details of these formulas are provided in Table S1. In the figure, the black triangles denote the reference values obtained from the side-capacitor measurements. It is observed that the curve associated with the discovered law closely follows the reference points over the entire range of $E$, with only small deviations at both low and high surface gradients. In the low-field region, almost all the empirical models systematically overestimate the measured levels. Around the mid-range of $E$, several empirical models track the data well, particularly the BPA model. However, most models still show an offset relative to the reference. At higher surface gradients, most empirical formulas, such as FGH and BPA, produce substantially steeper increases than those observed, leading to significant overestimation of audible noise. In contrast, the ENEL and IREQ models exhibit good consistency with the observation data. Overall, existing empirical models achieve good predictive performance only over limited ranges of the surface gradient, whereas the discovered law maintains consistently accurate behaviour across the entire range considered. Moreover, empirical models typically contain four or more terms and are restricted to specific log-linear structures. In contrast, the discovered formula uses only three terms yet outperforms the classical empirical laws over the entire experimental range. These findings illustrate the advantage of data-driven equation discovery over empirical modelling based solely on human intuition in finding a concise and accurate formula.

**Table 1. Comparison between the discovered audible noise predictive formula and existing empirical models for full-scale ultrahigh-voltage (UHV) transmission lines.** The unit of distance is *m*, and the unit of measured and predicted noise is dB. The differences between the predictions and observations are given in brackets. Bold entries indicate the best performance in each comparison.

| Lines | Distance | Measured | BPA | ENEL | EPRI | Ours |
|---|---|---|---|---|---|---|
| Project_UHV_8 | 15 | 50.2 | 51.7 (+1.5) | 51.7 (+1.5) | **51.0 (+0.8)** | 52.1 (+1.9) |
| Project_UHV_12 | 15 | 55.8 | 53.6 (-2.2) | 52.9 (-2.9) | 48.2 (-7.6) | **53.7 (-2.1)** |
| Project_UHV_16 | 15 | 50.8 | 51.9 (+1.1) | 51.6 (+0.8) | 43.2 (-7.6) | **51.4 (+0.6)** |
| HN-JZ | 15 | 45.2 | 46.7 (+1.5) | 47.4 (+2.2) | 46.0 (+0.8) | **46.0 (+0.8)** |
| HB-ZX | 10 | 46.7 | 47.2 (+0.5) | 47.9 (+2.2) | 46.5 (-0.2) | **46.5 (-0.2)** |
| HB-ZX | 20 | 46.9 | 47.8 (+0.9) | 48.4 (+1.5) | 47.1 (+0.2) | **47.1 (+0.2)** |
| HB-ZY | 10 | 47.8 | **47.9 (+0.1)** | 48.6 (+0.8) | 47.1 (-0.7) | 47.3 (-0.5) |
| HB-ZY | 20 | 47.0 | **47.3 (+0.3)** | 49.0 (+1.0) | 46.5 (-0.5) | 46.6 (-0.4) |
| **Average absolute error** | | | 1.01 | 1.61 | 2.30 | **0.84** |

We also compared the discovered equation with established symbolic regression baselines, including PySR[30] and deep symbolic optimization (DSO)[20]. The functional forms identified by these methods are reported in Table S1. PySR produces a relatively simple expression with limited nonlinearity, which is close in structure to conventional empirical formulations. Even with more terms than our model (four terms), PySR did not match our predictive performance (RMSE=1.152, MRE=0.362). In contrast, DSO identified a substantially more complex expression, comprising five terms and a rational structure that introduces stronger nonlinearity. Nevertheless, its predictive accuracy remained inferior (RMSE=2.112, MRE=0.540). Overall, compared with these baseline methods, our proposed equation achieves a more favourable trade-off between parsimony and predictive accuracy.

Afterwards, the performance of the discovered audible noise predictive formula is further evaluated on full-scale ultrahigh-voltage transmission lines outside the corona-cage environment. For each line, the law was first used to compute the A-weighted audible noise level associated with the local bundle configuration and surface gradient. These levels were then converted to ground-level sound pressure levels at specified observation distances using standard propagation and configuration corrections based on the actual line geometry. The complete transformation from bundle-level emission to field-measured noise is derived in detail in Supplementary Information S1.1.

Table 1 presents a comparative evaluation of the measured audible noise levels from UHV transmission lines in both China and America against predictions from three empirical models (BPA, ENEL, and EPRI) and our discovered law. In terms of the average mean error, the discovered law results in a significantly smaller overall error than the empirical benchmarks, indicating that the optimal performance is achieved under most of the tested conditions. With respect to the three UHV projects in America, which correspond to lines employing 8-, 12-, and 16-subconductor bundles, our law achieves the best performance in the 12- and 16-bundle configurations. The BPA model appears to systematically overestimate the measured levels, suggesting a biased formulation that reflects the limitations of manual empirical fitting. In contrast, the errors of our discovered equation do not exhibit a consistent directional bias across cases. Notably, although the law was originally derived from corona-cage data encompassing only 4 to 8 subconductors, it maintains high predictive accuracy when applied to configurations with a greater number of subconductors, effectively outperforming established empirical models. This demonstrates its strong generalization capability beyond its original training domain.

Then, we utilized field data from China's 1000 kV Jindongnan–Nanyang–Jingmen UHV AC line for comparison, with measurements collected at the Jiaozuo (HN-JZ), Zhongxiang (HB-ZX), and Zaoyang (HB-ZY) sites at distances of 10–20 m. The discovered law outperforms existing equations at both the HN-JZ and HB-ZX sites. Notably, while empirical models occasionally yield higher accuracy in specific instances (e.g., BPA for the HB-ZY case), they exhibit larger prediction errors in other scenarios, resulting in greater overall error. This inconsistency highlights the inherent limitations of traditional empirical models in that their effectiveness is usually constrained in limited scenarios. Conversely, the discovered law presents robust and consistent predictions across all the varied scenarios.

**Discovery of the radio-interference excitation function for AC lines**

Radio interference produced by corona discharges on high-voltage AC lines manifests as impulsive electromagnetic noise in the radio-frequency band and can severely affect nearby communication systems. In engineering practice, the radiated interference level at a given observation point is usually expressed as a radio-interference voltage or field strength, which depends not only on the intrinsic corona activity of the conductors but also on the line geometry, tower configuration and propagation conditions. To separate these effects, the concept of a radio-interference excitation function (RIEF) is typically employed[31]. The RIEF characterizes the intrinsic ability of a bundle conductor to generate radio interference under specified electrical conditions and is defined in decibels (dB). Once the RIEF is known as a function of the local surface gradient and conductor parameters, the radiated interference at ground level can be obtained by applying geometrical and propagation corrections that account for the phase arrangement, tower height and observation distance[32]. Details of the derivations are provided in Supplementary Information S1.2.

For the purpose of data-driven equation discovery, the RIEF was modelled as a function of the same three predictors used in the audible noise analysis: the surface gradient of the conductor bundle $E$, the number of subconductors in the bundle $n$, and the subconductor diameter $d$. The bundle configurations in the experiments are the same as those in the audible noise experiments. For each configuration, the corona was initiated and controlled by adjusting the applied voltage to cover a broad range of surface gradients. The RIEF value was quantified in the corona-cage tests using the side-capacitor method. In this approach, the interference voltage coupled through a measuring capacitor was recorded. This voltage was subsequently processed according to a standardized procedure to determine the equivalent RIEF value, $\Gamma_{RI}$, expressed in dB.

The graph-based equation-discovery framework was applied to this dataset under settings analogous to those used for audible noise. Candidate analytic forms were constructed from the operator set ($+, \times, pow, log$). The search was performed with a population size of 500 over 200 optimization iterations. To ensure physical consistency, a monotonicity penalty with a weight of 0.1 was integrated into the optimization objective. This regularization is grounded in the established engineering principle that $\Gamma_{RI}$ must increase monotonically with the surface gradient, bundle size, and conductor diameter. As in the audible noise case, this constraint was enforced over a feasible domain that extended beyond the experimental range, promoting physically reasonable extrapolation.

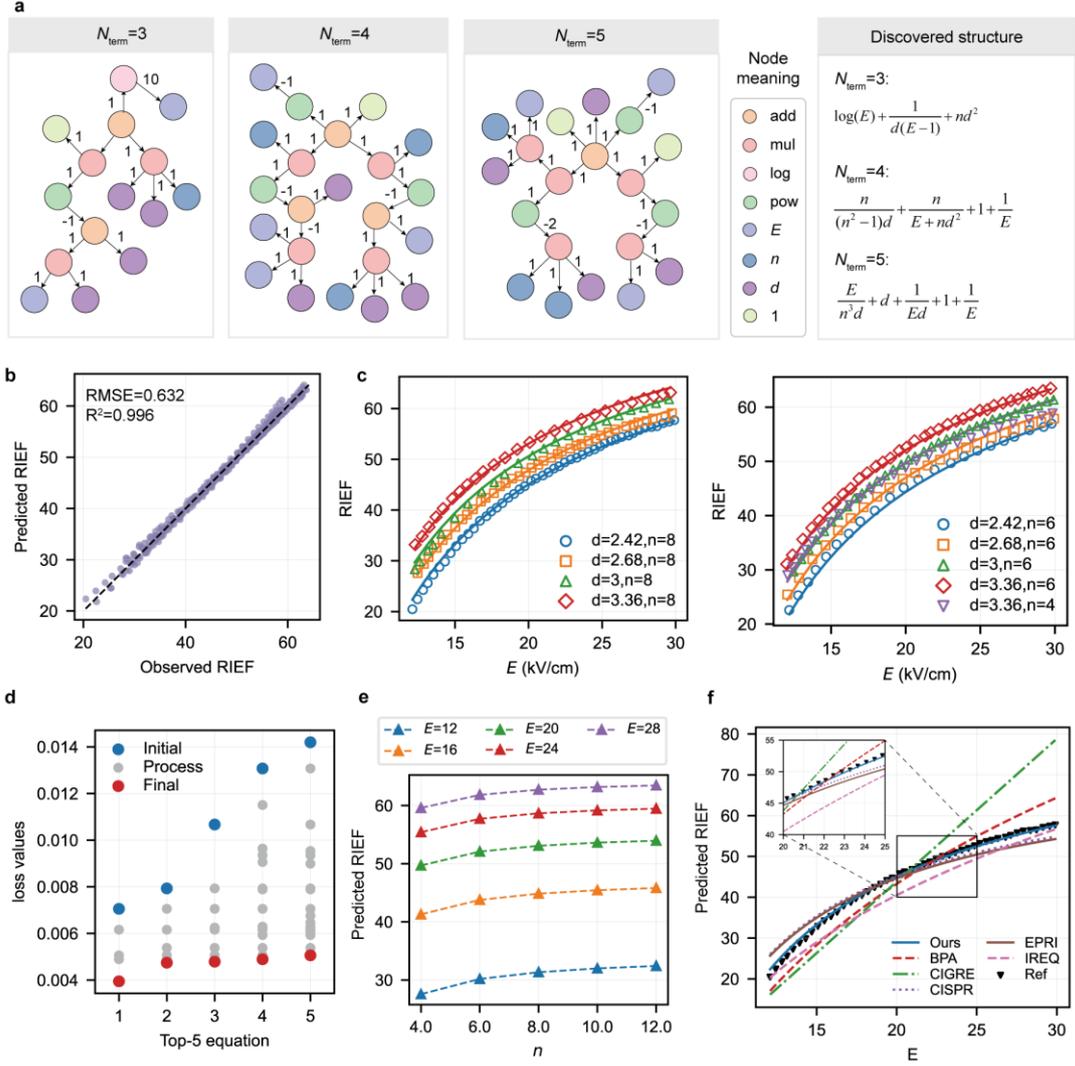

**Fig. 3. Results of data-driven equation discovery for the radio-interference excitation function $\Gamma_{RI}$. (a)** The discovered best graphs and formula structures under different numbers of maximum terms. **(b)** Comparison between the predicted and observed $\Gamma_{RI}$ across all data. **(c)** Comparison between the predicted and observed $\Gamma_{RI}$ for subconductor counts $n$=8, 6, and 4 with different subconductor diameters $d$. Here, the scatters represent the observed data, and the lines represent the predictions from the discovered formula. **(d)** Loss values of the top-5 equations during optimization. **(e)** Predicted $\Gamma_{RI}$ under different subconductor counts $n$ and surface gradients $E$. **(f)** Comparison between the discovered formula and classical empirical models for the 8×LGJ300/50 bundle conductor.

To explore the relationship between structural complexity and predictive performance, the maximum number of terms in the candidate expressions was limited to 3, 4 and 5 (Fig. 3a). The best-performing formulas under each constraint are as follows:

$$\Gamma_{RI}=45.6\log(E) - \frac{819.5}{d(E-1)} + 0.07nd^2 \tag{9}$$

$$\Gamma_{RI}= -\frac{117.2n}{n^2d\text{-}d} - \frac{133.5n}{E+nd^2}+98.68 - \frac{629.7}{E} \tag{10}$$

$$\Gamma_{RI}= -\frac{45.87E}{n^3d}+4.499d+72.88 - \frac{522.2}{E} - \frac{543.4}{Ed} \tag{11}$$

For comparison, a conventional polynomial-type regression inspired by classical empirical forms was also fitted:

$$\Gamma_{RI}=6.51d+10.287\log(n)+55.22 - \frac{671.7}{E} \tag{12}$$

On the full corona-cage dataset, the discovered formulas (Eqs. (9)–(11)) achieve RMSE values of 0.931, 0.632, and 0.521, with corresponding MREs of 0.017, 0.012, and 0.010, respectively. Benchmarking against a carefully tuned polynomial baseline (Eq. (12), RMSE = 1.021, MRE=0.019) reveals that even the simplest discovered expression reduces the RMSE by approximately 10%.

The algebraic forms of Eqs. (9)–(11) provide insight into how radio-interference excitation depends on the conductor geometry and surface gradient. The negative inverse term of $E$ in Eq. (10) and (11) captures the observed low-field suppression of radio interference; below the fully developed corona regime, modest increases in the surface gradient lead to a rapid increase in the RIEF value, whereas at higher gradients, the effect saturates, which is consistent with the $1/E$ dependence. When accuracy is balanced against interpretability and ease of application, the four-term law in Eq. (10) provides a favourable compromise. It improves the RMSE by approximately 0.39 compared with the polynomial baseline (Fig. 3b) while maintaining a relatively simple functional structure. Therefore, Eq. (10) is adopted as the preferred radio-interference excitation law for subsequent line-level predictions. Notably, the discovered RIEF laws retain the qualitative monotonic trends expected from corona physics (Fig. 3c). The monotonicity penalty used during optimization effectively prevents nonphysical decreases in the discovered RIEF formula with increasing corona severity (Fig. 3d). The monotonic behaviour of the discovered law is evident in Fig. 3e, where the predicted curves increase smoothly with the variable $n$ across the entire domain. Consequently, the data-driven expressions not only achieve high numerical accuracy but also exhibit behaviour consistent with the established understanding of radio-interference generation on AC transmission lines, providing a reliable basis for the subsequent prediction of radiated interference levels on practical UHV lines.

**Comparison between the discovered radio-interference excitation function and existing empirical models**

The performance of the discovered RIEF formula was first examined against the corona-cage measurements from which the RIEF data were generated. The predicted excitation function as a function of the surface gradient $E$ for a representative bundle configuration is shown in Fig. 3f. The black triangles denote the reference values obtained from the side-capacitor measurements, while the coloured curves correspond to the discovered law and to established empirical formulations, including the BPA, CIGRE, CISPR, EPRI and IREQ expressions. Details of these empirical models are provided in Table S2. Across the entire range of surface gradients from approximately 12.5 to 30 kV/cm, the curve associated with the discovered law closely tracks the reference data, maintaining a nearly parallel trend and very small absolute deviations. At low surface gradients, empirical models such as EPRI and CISPR exhibit relatively small errors, but their predictions still fall short of those obtained with the discovered law. In contrast, the CIGRE and BPA formulas tend to underestimate the measured values systematically. Interestingly, as the surface gradient increases to the mid-range (approximately 20 kV/cm), almost all the empirical formulas yield reasonably accurate predictions. However, at higher surface gradients, their behaviour begins to diverge. In this regime, CISPR and IREQ maintain relatively good agreement with the measurements, whereas other empirical models, such as CIGRE, exhibit substantially larger errors. In contrast, the discovered model presents consistently accurate predictions across the entire range of surface gradients, demonstrating superior stability and generalizability.

Similarly, we compared our discovered equation with established symbolic regression baselines, including PySR and DSO. The functional forms returned by these methods are reported in Table S2. For this task, PySR produced a highly complex expression, featuring nested logarithms to capture nonlinearity and additional multiplicative couplings among terms. Although its predictive performance is close to ours (RMSE=0.747, MRE=0.014), the resulting expression is substantially less parsimonious. In contrast, DSO yielded a much simpler form, consisting primarily of linear components and elementary rational terms, but with markedly worse accuracy (RMSE=3.302, MRE=0.061). Together, these results highlight a notable behaviour of existing baselines: the complexity of the discovered expressions can vary substantially across problems and is not consistently aligned with predictive performance. By comparison, our framework more stably balances parsimony and accuracy across tasks.

Beyond the corona-cage experiments, the discovered RIEF model was validated on full-scale UHV transmission lines. For line-level assessment, the RIEF serves as an intrinsic source parameter that is subsequently propagated along the line to obtain the radio-interference level at ground observation points. The propagation model treats each differential line segment as an equivalent noise source characterized by its local excitation function $\Gamma_n$. The contribution of each segment to the field at the receiver is modulated by frequency-dependent transfer functions and attenuation factors, and the total interference is obtained via superposition over all segments and modes. Details of this transformation are provided in Supplementary Information S1.2.

Using this procedure, predictions from the discovered law were compared against three widely used empirical formulations (BPA, HVTRC, and CISPR) using measurement data from four UHV AC lines located in the United States, Canada, and China. Table 2 summarizes the measured RI levels, model predictions, and corresponding errors. Overall, the discovered law demonstrates robust predictive accuracy for RI across UHV transmission lines in different countries, yielding a significantly smaller average absolute error compared with conventional empirical models. In terms of specific model performance, the HVTRC formula performs well for North American lines since it was originally developed using data from similar systems. However, when applied to UHV lines in China, its accuracy decreases and remains inferior to our discovered law. The BPA formula shows a consistent negative bias for all of these scenarios. Notably, BPA shows significant deviations when applied to Chinese UHV lines, which employ larger bundle configurations and operate at higher voltages. Conversely, while the CISPR model achieves higher accuracy for Chinese UHV lines, its performance degrades substantially when applied to the North American and Canadian scenarios. In contrast to the inconsistent performance of these empirical models, the discovered law performs comparably to or better than the best region-specific models in cases from different countries and avoids systematic biases.

**Table 2. Comparison between the discovered radio-interference excitation function and existing empirical models for full-scale ultrahigh-voltage (UHV) transmission lines.** The unit of distance is $m$. The differences between the predictions and observations are given in brackets. Bold entries indicate the best performance in each comparison.

| Lines | $n$ | Measured | BPA | HVTRC | CISPR | Ours |
|---|---|---|---|---|---|---|
| FT Wayne, USA (760 kV) | 4 | 70.5 | 69.9 (-0.6) | 71.2 (+0.7) | 74.4 (+3.9) | **70.1 (-0.4)** |
| Montreal, Canada (735 kV) | 4 | 68.0 | 66.0 (-2.0) | **68.5 (+0.5)** | 71.7 (+3.7) | 67.3 (-0.8) |
| Hubei, China (1050 kV) | 8 | 72.0 | 64.6 (-7.4) | 71.5 (-0.5) | 71.6 (-0.4) | **72.2 (+0.2)** |
| Anhui, China (1050 kV) | 8 | 71.3 | 63.5 (-7.8) | 70.4 (-0.9) | 70.5 (-0.8) | **71.1 (-0.2)** |
| **Average absolute error** | | | 4.45 | 0.65 | 2.20 | **0.40** |

Taken together, validation on both corona-cage data and full-scale UHV lines highlight several important features of the data-driven RIEF law. First, the law reproduces the detailed dependence of radio-interference excitation on the surface gradient with high fidelity, capturing both the low-field onset and the high-field saturation behaviour that traditional empirical formulas approximate only coarsely. Second, when combined with a physically based propagation model, the same law yields line-level RI predictions that are competitive with, and in several cases superior to, the best existing empirical models. Third, prior models often exhibit systematic bias, reflecting the fact that manually constructed equations are frequently tailored to a specific regime or class of operating conditions and therefore do not generalize well beyond that domain. In contrast, the discovered law can be generalized across systems spanning different voltage levels, bundle configurations and geographical regions.

**Discussion**

The present study demonstrates that corona-induced audible noise and radio interference on high-voltage AC transmission lines can be described by compact, data-driven discovered analytical laws that outperform long-standing empirical formulas while retaining full interpretability. From controlled corona-cage experiments on thirteen bundle configurations, a graph-based equation discovery framework was proposed to identify explicit relationships among the surface electric gradient, bundle geometry and two key descriptors of corona emissions: the A-weighted AN level and the RIEF. The discovered formula is able to reproduce the measured trends across a wide range of operating conditions and can be generalized successfully to full-scale UHV lines in different countries.

From a methodological perspective, the proposed graph-based equation discovery framework provides a new avenue for mathematical modelling in electrical engineering. Classical empirical approaches typically rely on human intuition to postulate candidate functional terms and then fit them using simple linear or log-linear regression. The resulting formulas are usually restricted to low-order polynomial structures and are therefore effective only within limited calibration ranges, with both predictive accuracy and generalizability constrained. Attempts to modify such empirical models by adding correction terms often lead to increasingly cumbersome expressions without resolving the underlying mismatch between the model structure and the true dependence[33]. At the other extreme, purely black-box machine-learning models can yield excellent numerical predictions, but they do not provide explicit insight into how electrical and geometric parameters shape corona emissions, nor do they guarantee physically reasonable extrapolation. In practical applications, their deployment also requires dedicated computational environments, which compares unfavourably with the simplicity and portability of closed-form formulas. The graph-based equation discovery framework adopted here provides an alternative pathway. By searching over explicit analytic expressions constructed from a small operator set, it preserves the transparency and ease of implementation of empirical laws while deriving their functional structure directly from data rather than from ad hoc assumptions. The results of this study show that, even when the algebraic complexity in terms of the number of terms is kept at or below that of traditional formulas, data-driven discovery can identify functional forms that more closely follow the measurements and maintain accuracy for configurations that lie beyond the historical calibration range.

The proposed graph-based equation discovery framework offers enhanced flexibility and more powerful structural search capabilities compared with conventional tree-based symbolic regression[34]. By generating and optimizing candidate equations through graph templates, the method maintains a diverse space of potential functional forms while incorporating mild structural priors. This approach effectively suppresses the emergence of unreasonable complex expressions that often plague unconstrained searches. A critically important feature of the proposed framework is the explicit integration of monotonicity constraints into the discovery process. In numerous engineering contexts, including corona phenomena, the qualitative dependencies between inputs and outputs are established a priori. Physical principles dictate that within operational ranges, audible noise and radio interference should increase monotonically with the surface electric field and generally with the effective conductor size. Conventional symbolic regression algorithms lack mechanisms to enforce such fundamental behaviour, frequently producing expressions that, despite fitting training data adequately, display nonphysical characteristics such as local oscillations or decreasing trends when extrapolated. By penalizing violations of prescribed monotonic relationships during optimization and evaluating candidates across extended feasible domains, our framework produces laws that achieve both accuracy on observed data and global consistency with physical trends. This represents an advancement for engineering applications where interpretability and robustness are important.

Several directions emerge for further development, together with some limitations of the present work. The analysis has been restricted to variables that can be measured and controlled reliably in corona-cage experiments, namely, the surface gradient and basic bundle geometry. As a result, the discovered laws are calibrated on a relatively narrow set of environmental and surface conditions, and their performance under extreme climates, complex terrain or heavily aged conductors has not been fully established. Although representative of typical UHV bundles, the corona-cage configurations do not cover the full variety of line designs in service worldwide, and field validation has thus far involved a limited number of UHV projects. In this sense, the monotonicity constraints and structural priors mitigate, but do not eliminate, the risk of bias when extrapolating beyond the regimes sampled in the present data. Moreover, the equation-discovery framework itself involves several hyperparameters that may influence the final formulas and could conceal alternative expressions with comparable accuracy but different interpretability. Overall, despite the above constraints, the results indicate that high-voltage transmission design can benefit substantially from moving beyond fixed empirical formulas towards data-driven discovered analytic laws discovered under explicit structural and physical constraints.

**Acknowledgements**

This work was supported and partially funded by the National Natural Science Foundation of China (Grant Nos. 12501744, 52288101, and 12572266), the Postdoctoral Fellowship Program and China Postdoctoral Science Foundation (Grant Nos. BX20250063 and 2024M761535), the National Key Research and Development Program (2024YFF1500600), and the Yongjiang Talent Program of Ningbo (2022A-242-G). This work was supported by the High Performance Computing Centers at Eastern Institute of Technology, Ningbo, and Ningbo Institute of Digital Twin.

**Author contributions**

H. X., C. K., Y. C. and D. Z. conceived the idea, designed the study, and analysed the results. H. X. developed the algorithm, performed the computations, and generated the results and figures. H. X., C. K., Y. C. and D. Z. wrote and edited the manuscript. C. K. and D. Z. supervised the entire project.

**Declaration of interests**

The authors declare that they have no competing interests.

**Data availability**

All datasets and codes generated in this work have been deposited in the open-source repository: https://github.com/woshixuhao/Discovery_of_corona/tree/main.

# Supplementary Information for Graph-based data-driven discovery of interpretable laws governing corona-induced noise and radio interference for high-voltage transmission lines

Hao Xu[1,2], Yuntian Chen[1,3,*], Chongqing Kang[2,*] and Dongxiao Zhang[1,4,*]

[1] Zhejiang Key Laboratory of Industrial Intelligence and Digital Twin, Eastern Institute of Technology, Ningbo, Zhejiang 315200, China

[2] Department of Electrical Engineering, Tsinghua University, Beijing 100084, P. R. China.

[3] Ningbo Institute of Digital Twin, Eastern Institute of Technology, Ningbo, Zhejiang 315200, P. R. China

[4] Institute for Advanced Study, Lingnan University, Tuen Mun, Hong Kong

[*] Corresponding authors.

Email address: ychen@eitech.edu.cn (Y. Chen); cqkang@tsinghua.edu.cn (C. Kang), dzhang@eitech.edu.cn (D. Zhang).

## 1. Supplementary text

### 1.1 Calculation from single-phase audible-noise generation to ground-level A-weighted noise

For comparison with field measurements, the audible-noise generation power of a single phase, expressed in terms of the bundle surface gradient and conductor geometry, must be transformed into the A-weighted sound-pressure level at a ground-level observation point. This subsection summarizes the procedure adopted in this study. For a bundle conductor with $n$ subconductors of equivalent diameter $d$ (cm) and average maximum surface gradient ($E$) (kV/cm), the A-weighted audible-noise generation power level per unit length of a single phase (also referred to as the sound-power level) is given as $L_{AN}$, which is the target of data-driven discovery for audible noise prediction model. To obtain the A-weighted sound-pressure level at a point on the ground, the contributions from all phases are superposed in the energy domain. For the $i^{\text{th}}$ phase, located at a horizontal distance and height that give a straight-line distance $R_i$ (m) from the observation point, the corresponding A-weighted sound-pressure level contribution is:

$$L_{p,i} = L_{AN,i} - C \log_{10} R_i - 5.8 \qquad (S.1)$$

where $C$=10 for the empirical models utilized in this paper and $C$=11.4 for the data-driven discovered formula. The term $\log_{10} R_i$ represents the cylindrical spreading of sound from a line source over a reflective ground plane. Assuming incoherent summation of the noise from different phases, the overall A-weighted sound-pressure level at the observation point is

$$L_p = 10 \log_{10}\left(\sum_{i=1}^{z} 10^{L_{p,i}/10}\right) \tag{S.2}$$

where $z$ is the number of phases. For a conventional three-phase line, $z=3$. For a typical three-phase transmission tower, the distances $R_i$ are obtained directly from the tower geometry. Denote by $(x_i, h_i)$ the horizontal coordinate and height of the center of the $i^{\text{th}}$ phase conductor relative to the tower centerline, and by $(x_m, h_m)$ the coordinates of the measurement point with $h_m \approx 1.5$ m for a standard microphone height. The straight-line distance from the observation point to phase $i$ is then:

$$R_i = \sqrt{(x_m - x_i)^2 + (h_m - h_i)^2} \tag{S.3}$$

For a common horizontally arranged three-phase tower, the three phase conductors lie approximately at the same height $h_i$ but with different lateral offsets $x_i$; for a triangular configuration, both $x_i$ and $h_i$ differ among phases. In either case, once $(x_i, h_i)$ are known from the tower design and the measurement position $(x_m, h_m)$ is specified, the distances $R_i$ are computed from Eq. (S.3) and substituted into Eq. (S.4). This yields the predicted A-weighted ground-level audible-noise level at the receiver location, which can be directly compared with field measurements.

## 1.2 Conversion of RIEF to radio-interference levels for AC transmission lines

The radio-interference excitation function (RIEF), denoted $\Gamma_{\text{RI}}$, characterises the ability of a bundle conductor to generate radio-frequency corona currents under specified operating and weather conditions. At the interference frequency $f_{\text{RI}}$, the multiconductor line is modelled in the frequency domain by the per-unit-length series-impedance and shunt-admittance matrices **Z** and **Y**. These incorporate conductor resistance and internal inductance with skin effect, as well as ground-return parameters obtained from the complex penetration depth. The voltage and current phasors **V** and **I** along the line satisfy the telegrapher equations with distributed current sources:

$$\begin{aligned} \frac{d^2 U}{dx^2} &= \mathbf{ZYV} \\ \frac{d^2 I}{dx^2} &= \mathbf{YZI} \end{aligned} \tag{S.4}$$

These coupled equations are decoupled by modal transformation. Let **M** and **N** be the right-eigenvector matrices of **ZY** and **YZ**, respectively, with common eigenvalue matrix $\lambda$:

$$\lambda = \mathbf{M}^{-1}\mathbf{ZYM} = \mathbf{N}^{-1}\mathbf{YZN} \tag{S.5}$$

The corona currents density vector is obtained as:

$$\mathbf{J}=\frac{\mathbf{C}}{2\pi\varepsilon_0}\Gamma \tag{S.6}$$

Transforming to modal quantities,

$$\mathbf{J}_m=\mathbf{N}^{-1}\mathbf{J} \tag{S.7}$$

Then, considering that

$$\mathbf{I}_m=\mathbf{g}_m\mathbf{J}_m \tag{S.8}$$

The corona-induced current phasors in each conductor can be written as:

$$\mathbf{I}=\frac{\mathbf{N}\mathbf{g}\mathbf{N}^{-1}\mathbf{C}^{-1}\Gamma}{2\pi\varepsilon_0} \tag{S.9}$$

Once the corona currents are known, the electromagnetic field at a ground-level observation point $(x,0)$ is obtained by superposition of the fields produced by each conductor. The horizontal magnetic-field component $H_x(x,0)$ is computed as

$$H_x(x,0)=\sum_{i=1}^{N}\frac{I_i}{2\pi}\left[\frac{h_i+P}{(x-x_i)^2+(h_i+P)^2}-\frac{h_i-P}{(x-x_i)^2+(h_i-P)^2}\right] \tag{S.10}$$

where $I_i$ is the complex current in conductor $i$, $(x_i,h_i)$ are its horizontal position and height, and $P$ is the complex penetration depth of the ground return. The dominant vertical electric-field component is then:

$$E_y(x,0)=Z_0 H_x(x,0) \tag{S.11}$$

with $Z_0\approx 120\pi\ \Omega$ the free-space wave impedance. For multiple bundles per phase, Eq. (S.10) includes all subconductors explicitly, or the bundle may be represented by equivalent conductors located at the bundle centroids with appropriately scaled currents. The radio-interference level at the observation point is finally expressed as an electric-field level in dB above $1\ \mu V/m$:

$$\mathrm{RI}(x)=20\log_{10}\left(\frac{|E_y(x,0)|}{1\ \mu V/m}\right)\quad [\mathrm{dB}\ \mu V/m] \tag{S.12}$$

In this way, the RIEF function can be consistently propagated through the multiconductor line model to yield predicted radio-interference levels at arbitrary observation points along the corridor.

## 2. Supplementary Tables and Figures

**Table S1. Formulas for the six conventional empirical models and two models discovered by symbolic regression baselines for predicting the A-weighted sound-pressure levels of the audible noise.** Here, $E$ is average maximum bundle surface gradient (kV/cm), $d$ is diameter of a subconductor (cm), $n$ is number of subconductors in a bundle, and $log$ denotes base-10 logarithm. Here, each model is written here in terms of the A-weighted audible-noise generation power per unit length, $L_{AN}$, expressed as a level in dB(A) re 1~µW/m.

| Model name | Formulas |
|---|---|
| BPA | $L_{AN} = 120 \log E + 55 \log d + 26.4 \log n - 128.4$ |
| ENEL | $L_{AN} = 85 \log E + 45 \log d + 18 \log n - 71$ |
| IREQ | $L_{AN} = 72 \log E + 45.81 \log d + 22.71 \log n - 57.6$ |
| FGH | $L_{AN} = 2E + 45 \log d + 18 \log n - 0.3$ |
| GE | $L_{AN} = -\dfrac{655}{E} + 44 \log d + 20 \log n + 67.9$ |
| EPRI | $L_{AN} = 120 \log E + 54 \log d + 24.8 - 126$ |
| PySR | $L_{AN} = 1.58nd - 2.97n + 55.6 - \dfrac{915}{E}$ |
| DSO | $L_{AN} = -2.65\dfrac{Ed}{n^2} + 62.8d - \dfrac{64.8d}{\log(E)} + 0.47n - 17$ |

**Table S2. Formulas for the five conventional empirical models and two models discovered by symbolic regression baselines for predicting the radio-interference excitation function, $\Gamma_{RI}$.** Here, $E$ is average maximum bundle surface gradient (kV/cm), $d$ is diameter of a subconductor (cm), $n$ is number of subconductors in a bundle, and *log* denotes base-10 logarithm.

| Model name | Formulas |
|---|---|
| BPA | $\Gamma_{RI}=120\log\left(\dfrac{E}{15}\right)+40\log\left(\dfrac{d}{4}\right)+37.02$ |
| CIGRE | $\Gamma_{RI}=3.5E+6d-40.69$ |
| EPRI | $\Gamma_{RI}=\begin{cases}-\dfrac{580}{E}+38\log\left(\dfrac{d}{3.8}\right)+81.1, & n\leq 8 \\ -\dfrac{580}{E}+38\log\left(\dfrac{d}{3.8}\right)+86.1, & n>8\end{cases}$ |
| CISPR | $\Gamma_{RI}=70-\dfrac{580}{E}+35\log d-10\log n$ |
| IREQ | $\Gamma_{RI}=-90.25+92.42\log E+43.03\log d-K(n)$ <br> $K(n)=\begin{cases}0, & n=1,\\ 3.7, & n=2,\\ 6, & n\geq 3.\end{cases}$ |
| PySR | $\Gamma_{RI}=0.51(d+158.407)\ln(\ln((n-6.35)E))-613$ |
| DSO | $\Gamma_{RI}=11.1\dfrac{E}{n}+18.2d-68.6\dfrac{d}{n}+0.99n-16.1$ |